\documentclass[conference]{IEEEtran}
\usepackage[letterpaper, left=0.6in, right=0.6in, top=0.65in, bottom=0.9in, includefoot]{geometry}

\usepackage{graphicx}
\usepackage{subcaption}
\usepackage{url}
\usepackage{array}
\usepackage[utf8]{inputenc} 
\usepackage[T1]{fontenc}
\usepackage{amsmath}
\usepackage{mathtools}
\usepackage{blkarray}
\usepackage{amsfonts}
\usepackage{amssymb}
\usepackage{amsthm}
\usepackage{enumerate}
\usepackage{cite}
\usepackage{lipsum}
\usepackage{mathtools}
\usepackage{cuted}
\usepackage{calc}
\usepackage{color}
\usepackage{epsfig}
\usepackage{setspace}
\usepackage{pstricks}
\usepackage{cancel}
\usepackage{multirow}
\usepackage{mathrsfs}
\usepackage{algorithm}
\usepackage{algpseudocode}
\usepackage{multirow}
\usepackage{diagbox}
\usepackage{multicol}
\usepackage{booktabs,makecell}
\usepackage{mathrsfs}

\usepackage{enumitem}

\usepackage{enumitem}
\newtheorem{defn}{Definition}
\newtheorem{thm}{{\cal T}heorem}

\newtheorem{remark}{Remark}

\newtheorem{example}{Example}

\newcommand{\W}{\mathcal{W}}

\newcommand{\F}{\mathbb{F}}
\newcommand{\M}{\mathcal{M}}

\input{epsf.sty}

\begin{document}
	\title{Secure Coded Distributed Computing}
	\author{
		\IEEEauthorblockN{Shanuja Sasi and Onur Günlü}
		\IEEEauthorblockA{ Information Theory and Security Laboratory (ITSL), Linköping University, Sweden \\
			E-mail: $\{$shanuja.sasi, onur.gunlu$\}$@liu.se}
	}
	\maketitle
	\begin{abstract}
		In this paper, we consider two critical aspects of security in the \textit{distributed computing (DC)} model: \textit{secure data shuffling} and \textit{secure coded computing}. It is imperative that any external entity overhearing the transmissions does not gain any information about the \textit{intermediate values (IVs)} exchanged during the shuffling phase of the DC model. Our approach ensures IV confidentiality during data shuffling. Moreover, each node in the system must be able to recover the IVs necessary for computing its output functions but must also remain oblivious to the IVs associated with output functions not assigned to it. We design secure DC methods and establish achievable limits on the tradeoffs between the communication and computation loads to contribute to the advancement of secure data processing in distributed systems.
	\end{abstract}
	
	\begin{IEEEkeywords}
			Coded distributed computing, MapReduce framework, information-theoretic security, secure computing.
	\end{IEEEkeywords}
	\vspace*{-0.25cm}
	\section{Introduction} \label{intro}
	In the realm of mobile applications demanding low-latency responsiveness, edge computing has emerged as a scorching topic due to its capability to provide high computation speeds and low latency. {\it Distributed computing (DC)} models, mainly focusing on Hadoop MapReduce, provide a framework for edge computing designs in the literature. In the MapReduce framework, the computing task undergoes three phases. In the Map phase, input files are distributed to edge nodes for local processing, outputting {\it intermediate values (IVs)}. During the Shuffle phase, IVs are exchanged among edge nodes. Once an edge node accumulates enough IVs, it proceeds to compute the output function in the Reduce phase. However, data shuffling in the Shuffle phase significantly impacts the latency of the output function computation. In \cite{LMA},  the authors used a methodology, called {\it coded distributed computing (CDC)},  to exploit coding in data shuffling to reduce the communication load by a factor of the computation load in the MapReduce framework. In  \cite{YYW},  the authors employed \textit{placement delivery array (PDA)} designs to create a coded computing scheme. Originally introduced as a solution for the coded caching problem, the concept of PDA has gained prominence in the literature, along with several studies exploring CDC and rate-limited communications \cite{PDAmain,YTC,SGR,SG}.
	
	Beyond the need for reducing the communication load during the data shuffling process, security emerges as a pivotal challenge in the realm of edge computing.  
	In the context of linear network coding based content delivery, there are primarily two levels of security for data confidentiality: {\it weak security (WS)} and {\it information-theoretic security (ITS)}. The studies in \cite{wsnc2,wsnc3} focused on WS in the context of data shuffling. Specifically, when an attacker cannot receive a sufficient number of coded packets from the data shuffling process, it becomes unable to decode and acquire any IVs. In contrast, our investigation centers around ITS, ensuring that no information related to the IVs of data shuffling is leaked to potential attackers.
	
	In this paper, we consider two aspects of security in the MapReduce framework: {\it secure data shuffling} and {\it secure coded computing}, inspired from the secretive coded caching problem studied in \cite{cc2,cc3,cc4,cc5}. For {\it secure data shuffling}, any external entity that overhears the transmissions during the shuffling phase must not get any information about the IVs. For {\it secure coded computing}, each node should be able to recover the IVs required for computing its output functions but must not gain any information about the IVs of the output functions it is not assigned. We establish achievable communication and computation tradeoffs for these security problems.
	
	\vspace*{-0.1cm}
	\section{Background and Preliminaries}
	\label{problem defintion}
		\vspace*{-0.1cm}
	\subsection{MapReduce Framework}
	We consider the DC model with  MapReduce framework \cite{LMA}. In this model, there are $K$ nodes indexed by $[K]$, representing the set $\{1,2, \ldots , K\}$. The task is to compute $K$ output functions  $\{\phi_q:  q \in [K]\}$  from $N$ distinct input files $\mathcal{W} =\{W_n : n\in [N]\}$, for some positive integer $N$. Each function $\phi_q,$ for $q \in [K]$, maps all $N$ input files, where each file has $w$ bits, into a stream of $b$ bits, i.e., we have 
	$\phi_q : \F_{2^w}^{N} \rightarrow \F_{2^b}. $
	Suppose, for every $q \in [K]$, there is a linear map function $g_{q} : \F_{2^w} \rightarrow \F_{2^t}$. Assume that $g_{q}(.)$ maps the input file $W_n$ into an {\it intermediate value (IV)} $v_{q,n} = g_{q}(W_n) \in \F_{2^t}$ of $t$ bits, for each $n \in [N]$. Similarly, for every $q \in [K]$, assume that there is a reduce function $h_q :  \F_{2^t}^{N} \rightarrow \F_{2^b}$ which maps all IVs into the output function $\phi_q = h_q(v_{q,1}, \ldots , v_{q,N} ) \in \F_{2^b}$ of $b$ bits. With that, the output function $\phi_q$, for each  $q \in [K]$, can  be equivalently described as 
	$\phi_q(\W) = h_q(v_{q,1}, \ldots , v_{q,N}  ) 
	= h_q(g_{q}(W_1), \ldots , g_{q}(W_{N}) ).$
	The  function computation is carried out in three phases:
	\begin{enumerate}[leftmargin=*,label=\arabic*.]
		\item {\bf Map Phase:} 
		Each node $k \!\in\! [K]$ stores  a subset of files  $\mathcal{M}_k  \!\subseteq\! \W$, and computes its IVs
		$\{v_{q,n} : q \!\in\! [K], W_n \! \in\!  \M_{k}, n \!\in\! [N]\}.$
		\item {\bf Shuffle Phase:}  Each node $k \in [K]$ is assigned to compute an output function $\phi_k$. The set of all IVs which each node $k$ does not have access to and needs to recover for computing the assigned output function is given
		by  
		$\{v_{k,n}: W_n \in \W \backslash \M_{k},n \in [N]\}.$
		Each  node $k$ creates a bit sequence ${\bf X}_k \in \{0,1\}^{l_k}$ using the IVs it has access to and sends it through a broadcast link to  all other nodes. 
		\item {\bf Reduce Phase:}  
		Receiving the sequence ${\{{\bf X}_j\}}_{j \in [K]\backslash k}$, each  node $k \in [K]$ decodes all the IVs required to compute  its output function, i.e., we have
		$H\left ({\{v_{k,n}\}}_{n\in [N]}|\M_k,{\{{\bf X}_j\}}_{j \in [K]\backslash k} \right )=0.$
	\end{enumerate}
	We next define the computation and communication loads for the DC problem. In \cite{LMA}, the computation load is defined as  the total number of files mapped across  $K$  nodes normalized by the total number of files. We generalize the definition as follows.
	\begin{defn}
		(Computation Load): Computation load $r$ is defined as the total number of bits associated with the files mapped across  $K$  nodes normalized by the total size of the files.
	\end{defn}
	\begin{defn}
		(Communication Load \cite{LMA}): The communication load $L$ is defined as the total number of bits transmitted by the $K$ nodes over the broadcast channel during the Shuffle phase normalized by the number of bits of all IVs.
	\end{defn}

	\vspace*{-0.3cm}
	\subsection{Placement Delivery Array}
	Yan et al. \cite{PDAmain} introduced the concept of PDA to represent coded caching schemes with the goal of reducing sub-packetization levels. Since then, several coded caching schemes based on the PDA concept have been reported.
	\begin{defn}  ({\bf Placement Delivery Array}\cite{PDAmain}):
		For positive integers $K, F, Z,$ and $S,$ an $F \times K$ array $P = [p_{f,k}]$ with $ f \in [F],$ and $ k \in [K]$ composed of a specific symbol $*$ and $S$ positive integers $[S],$ is called a $(K, F, Z, S)$ placement delivery array (PDA) if it satisfies the following conditions:
		\begin{itemize}
			\item {\it A1:} The symbol $*$ appears $Z$ times in each column;
			\item {\it A2:} Each integer occurs at least once in the array;
			\item {\it A3:} For any two distinct entries $p_{f_1,k_1}$ and $p_{f_2,k_2}, s=p_{f_1,k_1} = p_{f_2,k_2} $ is an integer only if
			\begin{enumerate}
				\item $f_1$ $\neq f_2$ and $ k_1$ $\neq k_2,$ i.e., they lie in distinct rows and distinct columns; and
				\item $p_{f_1,k_2} = p_{f_2,k_1} = *,$ i.e., the corresponding $2 \times 2$ sub-array formed by rows $f_1, f_2$ and columns $k_1, k_2$ must be either of the following forms
				$ \begin{pmatrix}
				s & *\\
				* & s
				\end{pmatrix} $or 
				$\begin{pmatrix}
				*& s\\
				s & *
				\end{pmatrix}.$\qed
			\end{enumerate} 
		\end{itemize}
		\label{def:PDA}
	\end{defn}
	\begin{example}
		\label{PDA example}
		Consider an $4 \times 5$ array $P_1$ as given below. It satisfies conditions \textit{A1, A2} and \textit{A3}. There are $2$ stars in each column and a total of $4$ integers in the array. Hence,  $P_1$ is a $(5,4,2,4)$ PDA.
		
		\noindent
		{\small
		\begin{equation}
		\label{A1}
		P_1 =
		\begin{blockarray}{ccccc}
		\begin{block}{(ccccc)}
		* & * & * & 1 &  2 \\
		* & 1 & 2 & * & * \\
		1 & * & 3 & * &  4\\
		2 & 3 & * & 4 &  *\\
		\end{block}
		\end{blockarray}. 
		\end{equation}}
	\vspace*{-0.8cm}
	\end{example}
	\section{Main Results}
	We first define secure data shuffling and secure coded computing in Definitions \ref{ssecure data shuffling} and \ref{ssecure computation}, respectively. 
	 A secure data shuffling scheme is provided in Theorem~\ref{thm1}, where we are interested in the secure delivery requirement as in Definition~\ref{ssecure data shuffling}. In Theorem~\ref{thm2}, we obtain a secure coded computing scheme, where we require that each node must be able to decode only the IVs corresponding to its assigned output functions and not be able to obtain any information about the remaining IVs. The proofs of Theorems \ref{thm1} and \ref{thm2} are provided in Sections \ref{proof thm1} and \ref{proof thm2}, respectively.
	\begin{defn}
		\label{ssecure data shuffling}
		(Secure Data Shuffling) Any eavesdropper that
		overhears the transmitted symbols during the shuffling phase
		must not obtain any information about the contents of the
		IVs. Therefore, we have
		$I\left ({\{v_{q,n}\}}_{q \in [K] ,n\in [N]} ; {\{{\bf X}_j\}}_{j \in [K]} \right ) =0$.
	\end{defn}
	\begin{defn}
		\label{ssecure computation}
		(Secure Coded Computing) For each $k\in [K]$, we require
		$I\left ({\{v_{q,n}\}}_{q \in [K] \backslash k,n\in [N]};\M_k,{\{{\bf X}_j\}}_{j \in [K]\backslash k} \right )=0.$
	\end{defn}
	\begin{thm}
		\label{thm1}
		Suppose that we are given a $(K, F,Z,S)$ PDA $P= [p_{f,k}]$ for $ f\in [F],k\in[K]$, and for some integers $ K,F,Z,$ and $S$, such that each integer appears more than once in the PDA $P$. There exists a secure data shuffling scheme for a DC model which consists of $K$  nodes, and $\eta F$ number of files, for some positive integer $\eta$.
		For the corresponding DC model, the computation load is $r=\frac{Z}{F}$ and the communication load achievable is given by
		
		\vspace*{-0.2cm}
		\noindent
		{\small
		\begin{align}
			\label{comm load}
		L= \frac{S}{KF}+\sum_{g=2}^{K}\frac{S_g}{KF (g-1)}
		\end{align}		\vspace*{-0.4cm}}
	
		\noindent where $S_g$ is the number of integers in $[S]$ which appear exactly $g$ times in the PDA $P$. 
	\end{thm}
\begin{remark}
	Without employing secure data shuffling, as in \cite{YTC}, the computation and communication loads achievable for a given PDA are the same as in Theorem \ref{thm1}. However, implementing secure data shuffling incurs an additional overhead in terms of storing secret keys at the nodes (as discussed in Section \ref{proof thm1}).
\end{remark}
	\begin{thm}
		\label{thm2}
		Suppose that we have a $(K, F,Z,S)$ PDA $P= [p_{f,k}]$ for $ f\in [F],k\in[K]$, and for some integers $ K,F,Z,$ and $S$, such that each integer appears more than once in the PDA $P$. A secure  coded computing scheme for a DC model which consists of $K$ nodes, and $\eta (F-Z)$ files, for some positive integer $\eta$ can be derived from this PDA.
		For the corresponding DC model, the computation load is given by $r=\frac{Z}{F-Z}$. Furthermore, the communication load achievable is given by 
		
		\vspace*{-0.2cm}
		\noindent
		{\small
			\begin{align}
			\label{comm load coded}
			L= \frac{S}{K(F-Z)}+\sum_{g=2}^{K}\frac{S_g}{K(F-Z) (g-1)}
			\end{align}\vspace*{-0.4cm}}
		
		\noindent where $S_g$ is the number of integers in $[S]$ which appear exactly $g$ times in the PDA $P$. 
	\end{thm}
\begin{remark}
	When considering a given PDA for secure coded computation, the computation and communication loads increase by a factor of $\frac{F}{F-Z}$ as compared to the non-secure scenario, as in \cite{YTC}. Moreover, as for secure data shuffling, an overhead arises from storing secret keys at the nodes (as discussed in Section \ref{proof thm2}).
\end{remark}
\noindent Now, we illustrate Theorem \ref{thm2} with the help of an example.
\begin{example}
	\label{exmp1}
	Consider the $(5, 4,  2,4)$ PDA $P_1$ of Example \ref{PDA example}. Consider a DC model where there are $N = 4$ input files $\{W_1^1,W_1^2,W_2^1,W_2^2\}$, each of size  $3$ bits and $Q = 5$ output functions $\{\phi_1,\phi_2,\phi_3,\phi_4,\phi_5\}$ to be computed. The files are divided into 2 batches $\{B_1,B_2\}$ such that each batch $B_m,$ for $m \in [2]$, has two files, $B_m=\{W_1^m,W_2^m\}$. They form the secret vector  $[W_1^m, W_2^m]^T$. Also, select two random variables $V_1^m,V_2^m$ for $m \in [2]$, uniformly and independently from the finite field $\F_{2^3}$. We form a  key vector  $ [V_1^m, V_2^m]^T$ for each $m \in [2]$.
	Consider the  following Cauchy matrix:	
		$ \noindent {\footnotesize	\mathbf{D}=
			\begin{bmatrix}
				1 &  6 & 2 & 4\\
				6 & 1 & 4 & 2\\
				2 & 4 & 1 & 6\\
				4 & 2 & 6 & 1
		\end{bmatrix}}$.
	We multiply this with the concatenation of the secret and  key vectors to generate the shares corresponding to the batch $B_m$, for $m \in [2]$, i.e., we have
	$[A^m_1, A^m_2, A^m_3, A^m_4]^T =\mathbf{D}.
	[W^m_1, W^m_2, V^m_1, V^m_2]^T$.
	The row index $f \in [4]$ in the PDA $P_1$ represents the share $A_{f}^m, \forall m \in [2]$, and the column index $k \in [5]$ represents the node $k$. There exists a $*$ in a row indexed by $f$ and column indexed by $k$ if and only if the  node $k$ has access to the shares $\{A_f^m: m\in [2]\}$, for each $f \in [4]$ and $k \in [5]$.
	
	Assign the output function $\phi_k$ to  the node $k \in [5]$. Let $S_k$ denote the set of all integers present in column $k$, for  $k\in [5]$.  For each $k \in [5],s \in S_k$ and $m \in [2]$, a secret key $T_{k,s}^m$ of size $\frac{t}{g_s-1}$ bits is generated uniformly and independently from $\mathbb{F}_{2^{\frac{t}{g_s-1}}}$, where $g_s$ is the number of occurrences of the integer $s$ in the PDA $P_1$.
	The shares and secret keys assigned to the nodes are:
	
	\vspace*{-0.1cm}
	\noindent	{\small \begin{align}
			\M_{1} &= \{A_1^m,A_2^m,T_{1,1}^m,T_{2,1}^m,T_{4,1}^m,T_{1,2}^m,T_{3,2}^m,T_{5,2}^m: m\in [2]\}, \nonumber\\
			\M_{2} &= \{A_1^m,A_3^m,T_{1,1}^m,T_{2,1}^m,T_{4,1}^m,T_{2,3}^m,T_{3,3}^m: m\in [2]\}, \nonumber\\
			\M_{3} &= \{A_1^m,A_4^m,T_{1,2}^m,T_{3,2}^m,T_{5,2}^m,T_{2,3}^m,T_{3,3}^m: m\in [2]\}, \nonumber\\
			\M_{4} &= \{A_2^m,A_3^m,T_{1,1}^m,T_{2,1}^m,T_{4,1}^m,T_{4,4}^m,T_{5,4}^m: m\in [2]\}, \nonumber\\
			\M_{5} &= \{A_2^m,A_4^m,T_{1,2}^m,T_{3,2}^m,T_{5,2}^m,T_{4,4}^m,T_{5,4}^m: m\in [2]\}.
	\end{align}\vspace*{-0.3cm}}

\vspace*{-0.2cm}
	\noindent Each  node $k$ computes the linear map functions referred as coded IVs in the set $\{c_{q,f}^m = g_q(A_f^m): q \in [5], A_f^m \in \M_{k}, f \in [4],m\in [2]\}$.  	
	Consider the first column, i.e. column with index $1$ of $P_1$. The set of all integers present in this column is $S_{1} = \{1,2\}$. 
	Consider the entry $s= 1 $ in $ S_{1}$. The other entries which are $1
	$ are in the columns indexed by $2$ and $4$. Hence, we partition the symbols in $c_{1,3}^m$ into $2$ packets, each of equal size, i.e., 
		$c_{1,3}^m=\{c_{1,3}^{m,2},c_{1,3}^{m,4} \}$, for each $m \in [2]$.
	Next for the entry $s= 2$ in $ S_{1}$, we partition $c_{1,4}^m$ into $2$ packets, since the other entries which are $2$ correspond to  the columns $3$, and $5$ i.e., we have
		$c_{1,4}^m=\{c_{1,4}^{m,3},c_{1,4}^{m,5} \}.$
	Similarly, for each column indexed by $k \in [5]$, 
	for entries  corresponding to $1$ and $2$ we partition the corresponding symbols into $2$ packets of equal sizes, while for entries $3$ and $4$, we partition the symbols into $1$ packet (which is the symbol itself), as shown below. 
	
	\noindent {\small
		\begin{align*}
			c_{2,2}^m= \{c_{2,2}^{m,1},c_{2,2}^{m,4} \},
			c_{2,4}^m= \{c_{2,4}^{m,3}\},
			c_{3,2}^m= \{c_{3,2}^{m,1},c_{3,2}^{m,5} \},
			c_{3,3}^m= \{c_{3,3}^{m,2}\}, \nonumber\\
			c_{4,1}^m= \{c_{4,1}^{m,1},c_{4,1}^{m,2} \},
			c_{4,4}^m= \{c_{4,4}^{m,5} \},
			c_{5,1}^m= \{c_{5,1}^{m,1},c_{5,1}^{m,3} \},
			c_{5,3}^m= \{c_{5,3}^{m,4} \}.
	\end{align*}\vspace*{-0.4cm}}

\noindent
	Since $|S_k|=2, \forall k\in [5]$, each node $k$ transmits two coded symbols $X_{k,s}^m, s\in S_{k}$ for each $m \in [2]$. The following are the coded symbols transmitted by the nodes:
	{\small \begin{align*}
			X_{1,1}^{m} &= c_{2,2}^{m,1}  \oplus c_{4,1}^{m,1} \oplus T_{1,1}^m\mathbf{,} \hspace{0.1cm}
			X_{1,2}^{m} = c_{3,2}^{m,1}  \oplus c_{5,1}^{m,1} \oplus T_{1,2}^m\mathbf{,} \nonumber\\
			X_{2,1}^{m} &= c_{1,3}^{m,2}  \oplus c_{4,1}^{m,2} \oplus T_{2,1}^m\mathbf{,} \hspace{0.1cm}
			X_{2,3}^{m} = c_{3,3}^{m,2} \oplus T_{2,3}^m\mathbf{,} \hspace{0.1cm}X_{3,3}^{m} = c_{2,4}^{m,3} \oplus T_{3,3}^m,  \nonumber\\
			X_{3,2}^{m} &= c_{1,4}^{m,3}  \oplus c_{5,1}^{m,3} \oplus T_{3,2}^m\mathbf{,} \hspace{0.1cm}
			X_{4,1}^{m} = c_{1,3}^{m,4}  \oplus c_{2,2}^{m,4} \oplus T_{4,1}^m\mathbf{,}  \nonumber\\
			X_{4,4}^{m} &= c_{5,3}^{m,4} \oplus T_{4,4}^m\mathbf{,}\hspace{0.1cm}
			X_{5,1}^{m} = c_{1,4}^{m,5}  \oplus c_{3,2}^{m,5} \oplus T_{5,1}^m\mathbf{,} \hspace{0.1cm}
			X_{5,3}^{m} = c_{4,4}^{m,4} \oplus T_{5,3}^m.
	\end{align*}}
\vspace*{-0.4cm}

	The  node $1$ can retrieve $c_{1,3}^{m,2}$  from the coded symbol $X_{2,1}^{m}$ transmitted by  node $2$, since it can compute $c_{4,1}^{m,2}$ from the shares in $\M_{1}$ and cancel out $T_{2,1}^m$. Similarly, it can retrieve $c_{1,3}^{m,4},c_{1,4}^{m,3}$ and  $c_{1,4}^{m,5}$ as well.
	Hence,   node $1$ can compute ${\bf D}^{-1}[c_{1,1}^{m}, c_{1,2}^{m}, c_{1,3}^{m}, c_{1,4}^{m}]^T$ for each $m\in [2]$ and retrieve the IVs required for computing the output function $\phi_k$. Also, it does not obtain any additional coded IVs not related to output function $\phi_k$. Given the access structure of the $(2,4)$ non-perfect secret-sharing scheme, having $2$ shares related to other output functions (hence, two coded IVs related to other output functions) maintains the secrecy.
	
	Each node $k$ stores $4$ shares $\{A_f^m\!:\! m\!\in\! [2], p_{f,k}=*, f\in [4]\}$, as in (\ref{shares}). Thus, the computation load is $\frac{4*w}{4*w}=1.$ In total $10$ coded symbols are transmitted across the nodes. The symbols corresponding to the entries $1$ and $2$ in the PDA $P_1$ are of size $\frac{t}{2}$ bits, whereas  symbols corresponding to the entries $3$ and $4$ are of size $t$ bits.  Thus, the communication load is $L =\frac{(\frac{t}{2} *6+t*4)*2}{5*4*t} =0.7$. For non-secure computing, the computation and communication loads are $0.5$ and $0.35$, respectively \cite{YTC}.\qed
\end{example}

\vspace*{-0.3cm}
	\section{Proof of Theorem \ref{thm1}}
	\label{proof thm1}
	We present the proof of Theorem \ref{thm1} in this section. 
	First, we assume that there are $\eta F$ number of files, $\{W_f^m:f \in [F],m\in [\eta]\}$, for some positive integer $\eta$. The $\eta F$ files are divided by grouping the files into $F$ disjoint batches $\{B_{1},B_{2},\ldots, B_{F}\}$ each containing $\eta$ files such that $ B_{f} = \{W_f^m: m \in [\eta]\}$, for $f \in [F]$. 
	
	\vspace*{-0.2cm}
	\subsection{Map Phase of Theorem \ref{thm1}}
	\label{map phase}
	We define $S_k$ as the set of all integers present in the column indexed by $k\in [K]$ and $g_s$ as the number of occurrences of the integer $s\in [S]$ in the PDA. The nodes are filled  as follows:
	\begin{itemize}
		\item Node $k \in [K]$ stores  all the files from the $f^{\text{th}}$ batch if the entry corresponding to the row indexed by $f$ and the column indexed by $k$ in the PDA $P$ is the symbol $``*"$, i.e., if $p_{f,k} =*$, for $f \in [F]$.
		\item For each $k \in [K],s \in S_k$ and $m \in [\eta]$, a random variable $T_{k,s}^m$ of size $\frac{t}{g_s-1}$ bits, referred to as a secret key, is generated uniformly from $\mathbb{F}_{2^{\frac{t}{g_s-1}}}$ such that the keys are independent of each other and the map functions.
		\item If $s$ appears in the column indexed by $k$ of $P$, i.e., if $p_{f,k} = s$ for some $f \in [F]$, then the secret key $T_{\hat{k},s}^m, \forall m \in [\eta],$ is stored at node $k$, if $p_{\hat{f},\hat{k}} = s$ for some $\hat{f} \in [F] ,\hat{k} \in [K]$.
	\end{itemize}
	 The subscript $s$ indicates the transmission instance during which the corresponding key is utilized to encrypt messages during the shuffling phase. The content stored at node $k \in [K]$ is:
	\vspace*{-0.1cm}
	
	\noindent
	{\small
	\begin{align}
	\label{store}
	\M_k = \Bigg\{\bigcup_{\substack{f \in [F] : \\p_{f,k}=*, m \in [\eta]}}W_f^m \Bigg\} \bigcup \Bigg\{\bigcup_{\substack{\hat{k} \in [K],\hat{f} \in [F] :\\ s \in S_k, p_{\hat{f},\hat{k}}=s, m \in [\eta]}}T_{\hat{k},s}^m \Bigg\}.
	\end{align}\vspace*{-0.3cm}}
	
	\noindent Each node $k$ computes all the map functions for the files in the set $\M_k$, i.e., it computes $v_{q,f}^m =g_{q}(W_f^m)$, for  $q \in [K],$ and $W_f^m \in \M_k,$ where $f \in [F],$ and $m \in [\eta]$.
	\subsection{Shuffle Phase of Theorem \ref{thm1}}
	\label{shuffle}
	Each node $k\in [K]$ is responsible for computing an output function $\phi_k$. The set of all IVs related to the output function $\phi_k$ it can compute using the accessible files is  $\{v_{k,f}^m\!:p_{f,k} \!=\!*,f \in [F],m \in [\eta]\}$. To complete the computation, node $k$ need the remaining IVs, represented by the set $\{v_{k,f}^m:p_{f,k} \neq*,f \in [F],m \in [\eta]\}$. These IVs are necessary for computing $\phi_k$.
	
	Consider each pair $(f, k)$ where $ f\in [F] $ and $ k \in[K]$. Suppose  $p_{f,k} = s $ for some $s\in [S]$. Let $g_s$ denote the total number of occurrences of $s$. Assume that the remaining $g_s-1$ occurrences of $s$ are distributed across   $\{(f_i,k_i): i\in [g_s-1]\}$ such that $
	p_{f_1,k_1} = p_{f_2,k_2} = \ldots = p_{f_{g_s-1},k_{g_s-1}} = s. $ Importantly, for each $k_i \in \{k_1,k_2,\ldots,k_{g_s-1}\}$ we know that $p_{f,k_i} =* $ (since $f \neq f_i$), as indicated by condition A3-2). 
	We partition the symbols in $v_{k,f}^m$ into $g_s-1$ packets each of equal size, for each $m \in [\eta]$, i.e., 
	
	\noindent
	{\small
	\begin{align}
	\label{partition symbols}
	v_{k,f}^m=\{v_{k,f}^{m,k_1}, v_{k,f}^{m,k_2}, \ldots, v_{k,f}^{m,k_{g_s-1}}\},&& \forall m \in [\eta].
	\end{align}}
	\noindent The shuffling phase consists of $S$ transmission instances, where each transmission instance is denoted by $s\in [S]$. During the transmission instance $s \in S$, we consider a set of nodes $U_s = \{k: p_{f,k} =s, k\in [K], f\in [F]\}$. For each $k \in U_s$, node $k$ multicasts the following messages of length $\frac{t}{g_s-1}$ bits each.
	\vspace*{-0.1cm}
	
	 \noindent {\small
	\begin{align}
	\label{transmit}
	X_{k,s}^m = \Bigg(\bigoplus_{\substack{(u,e)\in [F] \times ([K] / k) :\\ p_{u,e}=s}} v_{e,u}^{m,k} \Bigg)\bigoplus  T_{k,s}^m, && \forall  m \in [\eta].
	\end{align}\vspace*{-0.0cm}}
	Thus, the  messages transmitted by node $k$ can be expressed as
	${\bf X}_{k} = \cup_{ s \in [S]:p_{f,k}=s, f\in [F], m \in [\eta]} X_{k,s}^m.$
	The security of message delivery is ensured by XOR-ing each message with a secret key. Thus, external eavesdroppers attempting to wiretap the shared link are thwarted. These eavesdroppers remain uninformed about the IVs since they lack access to the uniformly distributed keys. 
	The node $k$ can create the message $X_{k,s}^m$ from the IVs accessible to it. In fact, for each $(u,e)$ in the sum (\ref{transmit}), there exists a corresponding $f \in [F]$ such that $p_{u,e} = p_{f,k} = s$. Since $e \neq k$, we deduce that $u$ $ \neq f$ and $ p_{u,k} = *$ according to condition A3-2). Consequently, the node  $k$ can compute the IVs within the set  $\{v_{e,u}^m: m\in [\eta]\}$.
	\subsection{Reduce Phase of Theorem \ref{thm1}}
		During the reduce phase, the  node $k$  computes the output function $\phi_k$. 
	Upon receiving the messages $\{{\bf X}_j\}_{j \in [K]\backslash k}$, each  node $k$  decodes the IVs $\{v_{k,f}^m : f \in [F],m\in [\eta]\}$, with the help of the secret keys and the IVs it can compute. 
Specifically, it needs to determine the set of  IVs $\{v_{k,f}^m : W_f^m \in \W \backslash \M_k,f \in [F],m\in [\eta] \}$.
	   Without loss of generality, assume that $p_{f,k} = s \in S_k$. For each $k_i \in \{k_1,k_2,\ldots,k_{g_s-1}\}$, as defined in (\ref{partition symbols}), node $k$ retrieves the symbol $v_{k,f}^{m,k_i}$ from the message $X_{k_i,s}^{m}$ transmitted by the node $k_i$ for each $m \in [\eta]$ i.e., we have
	
	\noindent
{\small	\begin{align}
	\label{node l transmission}
	X_{k_i,s}^m = \Bigg(\bigoplus_{\substack{(u,e)\in [F] \times ([K] / k_i) :\\ p_{u,e}=s}} v_{e,u}^{m,k_i} \Bigg)\bigoplus  T_{k_i,s}^m.
	\end{align}\vspace*{-0.0cm}}
	The node $k$ stores secret keys denoted as $T_{k_i,s}^m$, where $m \in [\eta]$. Hence, it can cancel out $T_{k_i,s}^m$ from (\ref{node l transmission}). In (\ref{node l transmission}), for $e$ $ \neq k$, $p_{u,e} = p_{f,k}=s$ implies that $p_{u,k} = *$ by A3-2). Hence,   node $k$ can compute $v_{e,u}^{m,k_i}$. For $e = k$, $p_{u,e} = p_{f,k} = s$ implies $u = f$ by A3-1. Therefore, the  node $k$ can retrieve the symbol $v_{k,f}^{m,k_i}$ from the message in (\ref{node l transmission}) by canceling out the rest of the symbols. By collecting all the symbols $v_{k,f}^{m,k_i}$ in (\ref{partition symbols}),   node $k$ can compute $\phi_k$.
	
	Now, we evaluate the computation and communication load for this scheme. Each node stores $\eta Zw$ bits corresponding to the files $\{W_{f}^m:p_{f,k}=*,f\in[F],m\in[\eta]\}$ in (\ref{store}). Hence, the computation load is $r=\frac{\eta Zw}{\eta Fw}$ =$\frac{Z}{F}$.
	For each $s \in [S]$ occurring $g_s$ times, there are $\eta g_s$ associated messages sent, each of size $\frac{t}{(g_s-1)}$ bits by (\ref{transmit}). Let $S_g $ denote the number of integers which appear exactly  $g$ times in the array. 
	The communication load is given by
	
	\noindent
	{\small	\begin{align}
	L &= \frac{1}{K\eta Ft}\sum_{s=1}^{S} \frac{ g_s \eta t}{(g_s-1)} 
	= \frac{1}{KF }\sum_{g=2}^{K} \frac{gS_g}{(g-1)} \nonumber\\
	&=  \frac{\sum_{g=2}^{K} S_g}{KF}+\sum_{g=2}^{K} \frac{S_g}{KF (g-1)} 
	=  \frac{S}{KF}+\sum_{g=2}^{K} \frac{S_g}{KF (g-1)}. 
	\label{calculate}
	\end{align}}

\vspace*{-0.4cm}
	\section{Proof of Theorem \ref{thm2}}
	\label{proof thm2}
	Assume that there are $\eta (F-Z)$ number of files, $\{W_n^m:n \in [F-Z],m\in [\eta]\}$, for some positive integer $\eta$. The $\eta(F-Z)$ files are divided by grouping the files into $\eta$ disjoint batches $\{B_{1},B_{2},\ldots, B_{\eta}\}$ each containing $(F-Z)$ files such that $ B_{m} = \{W_n^m: n \in [F-Z]\}$. 
	
	We employ non-perfect secret sharing schemes \cite{Sec2} to encode the files in each batch. These schemes are designed such that accessing a subset of shares does not provide significant information about the secret, which in this case is the files in a batch. Only if all shares are combined can the original files be reconstructed. The non-perfect secret sharing scheme is next defined.
	\begin{defn}
		For each batch $B_m, m \in [\eta]$, with size $w(F-Z)$ bits, a $(Z, F)$ non-perfect secret sharing scheme generates $F$
		shares, $A_1^m, A_2^m,\dots, A_F^m,$ such that accessing any $Z$ shares
		does not reveal any information about the batch $B_m$, i.e.,
		$I(B_m; \mathcal{A}) = 0, \forall \mathcal{A} \subseteq \{A_1^m, A_2^m,\dots, A_F^m\},|\mathcal{A}| \leq Z.$
		Furthermore, the knowledge of $F$ shares is sufficient to
		reconstruct the secret (batch), i.e.,
		$H(B_m|A_1^m, A_2^m,\dots, A_F^m) = 0.$
	\end{defn}
In our scenario, a $(Z,F)$ non-perfect secret sharing scheme has been identified, where shares are of size $\frac{1}{F-Z}$ times the size of the secret ($w$ bits) \cite{Sec2}. 
	 In contrast to perfect secret sharing schemes \cite{Sec4}, which allocate shares of size equal to the secret size ($w(F-Z)$ bits), non-perfect schemes are more efficient in terms of computation and communication load.
	An example of non-perfect secret sharing schemes mentioned in the literature are ramp threshold secret sharing schemes, as described in \cite{Sec2}. 
	
	For each batch $B_m,m \in [\eta]$, the $(F-Z)$ files corresponding to $B_m$ arranged in a column forms the secret vector, which is a $(F-Z) \times 1$ column vector ${\bf W}^m := [W_1^m ,W_2^m ,\dots,W_{(F-Z)}^m]^T$, each element of which belongs to $\F_{2^w}$. We also select $Z$ random variables uniformly and independently from the finite field $\F_{2^w}$ to form the  key vector ${\bf V}^m := [V_1^m, V_2^m, \dots , V_Z^m]^T$ of dimension $Z \times1$. Let the share vector, corresponding to $B_m$ be a $F \times 1$ column vector ${\bf A}^m= [A_1^m,A_2^m,\dots, A_F^m]^T$, where $
	A_f^m \in \F_{2^w}, \forall f \in [F]$. 
	Define the linear mapping $\Pi$ as the transformation that maps the secret vector ${\bf W}^m$ to the share vector ${\bf A}^m$ corresponding to batch $B_m$. This mapping is represented as 
	$\Pi : \mathbb{F}_{2^w}^{F-Z} \times \mathbb{F}_{2^w}^{Z} \rightarrow \mathbb{F}_{2^w}^{F}$
	such that ${\bf A}^m = \Pi({\bf W}^m,{\bf V}^m)$ satisfies the following conditions:
		(i) $H({\bf W}^m|{\bf A}^m) =0$ (Correctness) and
		(ii) $H({\bf W}^m|\mathcal{A}) = H({\bf W}^m), \mathcal{A} \subset \{A_1^m,A_2^m,\dots, A_F^m\}; |\mathcal{A}| \leq Z$ (secrecy).
To implement the linear mapping as described above, an $F \times F$ Cauchy matrix ${\bf D}$ \cite{Cauchy} is utilized, operating in the finite field $\mathbb{F}_{2^z}$, where $z \geq 1 + \log_2F$. See \cite{LSS2} for a similar scheme.

Using a Cauchy matrix facilitates the generation of the share vectors ${\bf A}^m$ from the secret vectors ${\bf W}^m$ in a manner consistent with the conditions of the non-perfect secret sharing scheme. Condition (i) stipulates that $F$ shares are adequate for the recovery of the secret vector ${\bf W}^m$, while (ii) ensures that any subset of $Z$ or fewer shares does not disclose any information about the batch. For each batch $B_m$, the  key vector ${\bf V}^m$ is concatenated below ${\bf W}^m$ to form the vector ${\bf Y}^m =[{\bf W}^m;{\bf V}^m]$ of dimension $F \times 1$. Then, the Cauchy matrix ${\bf D}$ is multiplied with ${\bf Y}^m$ over $\mathbb{F}_{2^z}$ to obtain the share vector ${\bf A}^m$, expressed as
$	{\bf A}^m_{F \times 1} = {\bf D}_{F \times F}. {\bf Y}^m_{F \times 1}.$
	Therefore, for each $f \in [F]$, a share $A_f^m$ is computed as
	
	\noindent
	{\small
	\begin{align}
	A_f^m \!=\! \sum_{j \in [F]} d_{f,j} Y_j^m\! =\! \sum_{j=1}^{F-Z} d_{f,j} W_j^m \!+\!\sum_{j = F-Z+1}^{F} d_{f,j} V_{j-(F-Z)}^m. 
	\end{align}}
	Here, each share $A_f^m$ is a linear combination of the secret vector and  key vector, with coefficients derived from the Cauchy matrix. Let $v_{q,j}^m =g_{q}(W_j^m)$ be the IV obtained by mapping the input file $W_j^m$ using the linear map function $g_q(.)$, for  $q \in [K] ,j \in [F-Z],$ and $m \in [\eta]$. Also, let $c_{q,f}^m =g_{q}(A_f^m)$ be linear map function values referred as {\it coded IV}, for each $q \in [K]$ and $A_f^m,$ where $f \in [F],$ and $m \in [\eta]$. Then, we have
	
	\noindent
	{\small
	\begin{align}
	c_{q,f}^m &=g_{q}(A_f^m) \!=\! \sum_{j=1}^{F-Z} d_{f,j} v_{q,j}^m \!+\!\sum_{j = F-Z+1}^{F} d_{f,j}  g_{q}(V_{j-(F-Z)}^m). 
	\label{eq1}
	\end{align}}
	In other words, each coded IV $c_{q,f}^m$ is computed based on the corresponding share $A_f^m$, utilizing the mapping function $g_q(.)$. 
	For each  $q \in [K] $ and $m \in [\eta]$,  $F$ coded IVs corresponding to $\{c_{q,f}^m: f\in[F]\}$ arranged in a column forms the coded IV vector, which is a $F \times 1$ column vector, ${\bf C}_q^m := [c_{q,1}^m ,c_{q,2}^m ,\dots,c_{q,F}^m]^T$, where each element of which belongs to $\F_{2^t}$. Similarly, the map function values $\{v_{q,f}^m: f\in [F-Z] \} \cup \{g_{q}(V_{f}^m): f \in [Z]\}$  forms a $F \times 1$ column vector, ${\bf g}_q^m := [v_{q,1}^m ,\dots,v_{q,F-Z}^m,g_{q}(V_{1}^m),\ldots,g_{q}(V_{Z}^m)]^T$, where each element of which belongs to $\F_{2^t}$. Hence, we have ${\bf C}_q^m=\mathbf{D}.  {\bf g}_q^m$.
	Once the coded IV vector is available, node $k$ can compute $\mathbf{D}^{-1}{\bf C}_k^m$ for each $m \in [\eta]$. Notably, $\mathbf{D}$ is known to all nodes, and since a Cauchy matrix is full rank, $\mathbf{D}^{-1}$ always exists. Consequently, node $k$ can retrieve $\{v_{k,f}^m: f\in [F-Z] \}$, where the first $F-Z$ elements correspond to these IVs. With these IVs, node $k$ can then proceed to compute the output function. 
	
	\vspace*{-0.1cm}
	\subsection{Map, Shuffle and Reduce Phases of Theorem \ref{thm2}}
   The map phase is similar to that in Section \ref{map phase} except that each node $k$ stores a subset of shares instead of files as follows:
   
	\noindent
{\small	\begin{align}
	\label{shares}
	\M_k = \Bigg\{\bigcup_{\substack{f \in [F] : \\p_{f,k}=*, m \in [\eta]}}A_f^m \Bigg\} \bigcup \Bigg\{\bigcup_{\substack{\hat{k} \in [K],\hat{f} \in [F] :\\ s \in S_k, p_{\hat{f},\hat{k}}=s, m \in [\eta]}}T_{\hat{k},s}^m \Bigg\}.
	\end{align}\vspace*{-0.4cm}}
	
	\noindent Each node $k$ computes the map functions of the shares stored in $\mathcal{M}_k$, specifically, it calculates the coded IV $c_{q,f}^m = g_{q}(A_f^m)$ for each $q \in [K]$, and $A_f^m \in \mathcal{M}_k$, where $f \in [F]$ and $m \in [\eta]$. 
	
	The shuffling phase is also similar to that in Section \ref{shuffle} except that instead of exchanging IVs, coded IVs are exchanged. As each message is XOR’ed with a secret key, the delivery is secure against external eavesdroppers wiretapping on the shared link. Thus, eavesdroppers do not obtain any information about the coded IVs as they do not have access to the secret keys.

	In the reduce phase, each  node $k$ first retrieves all the coded IVs $\{c_{k,f}^m: f \in [F], m\in [\eta]\}$ using the secret keys and the coded IVs it can compute. From these coded IVs, it decodes the set of IVs $\{v_{k,f}^m : f \in [F-Z], m\in [\eta]\}$ using the Cauchy matrix ${\bf D}$, and finally computes the output function assigned to it. Each node stores $\eta Zw$ bits corresponding to the shares $\{A_{f}^m:p_{f,k}=*,f\in[F],m\in[\eta]\}$ in (\ref{shares}). Hence, the computation load is $r=\frac{\eta Zw}{\eta (F-Z)w}$ =$\frac{Z}{F-Z}$.
The communication load is calculated similar to (\ref{calculate}) and is given by

	\noindent
{\small	\begin{align*}
	L &\!=\! \frac{1}{K\eta (F-Z)t}\sum_{s=0}^{S-1} \frac{ g_s \eta t}{(g_s-1)} 
	\!=\!  \frac{1}{K(F-Z)}\left (S\!+\!\sum_{g=2}^{K} \frac{S_g}{ (g-1)}\right ) 
	\end{align*}}
where $S_g $ denotes the number of integers which appear exactly  $g$ times in the array. 

	\subsection{Proof of Secrecy of Theorem \ref{thm2}}  
	To demonstrate that nodes gain no information about the content of any IVs corresponding to the output functions not assigned to them, we first establish that node $k$, for $k\in[K]$, cannot obtain any information about $\{v_{q,f}^m : q \in [K]\backslash k, f \in [F-Z],m \in [\eta]\}$ from the coded IVs it can compute. Denote an arbitrary $Z$-sized subset of ${\bf C}_q^m$ as $c^m_{q,i_1},c^m_{q,i_2},\ldots,c^m_{q,i_Z}$. We have:
	
\noindent	{\small \begin{align}
	\begin{bmatrix}
	c^m_{q,i_1}\\c^m_{q,i_2}\\ \vdots\\c^m_{q,i_Z}
	\end{bmatrix}
	&=
	\begin{bmatrix}
	d_{i_1,1}&d_{i_1,2}&\ldots&d_{i_1,F}\\
	d_{i_2,1}&d_{i_2,2}&\ldots&d_{i_2,F}\\
	\vdots& \vdots & \ddots & \vdots\\
	d_{i_Z,1}&d_{i_Z,2}&\ldots&d_{i_Z,F}\\
	\end{bmatrix} .  {\bf g}_q^m.
    \label{eq}
	\end{align}}
RHS of Eq. (\ref{eq}) can be written as 	$\mathbf{D_1}.[v_{q,1}^m, \ldots, v_{q,F-Z}^m]^T +$ $ \mathbf{D_2}. [g_{q}(V_{1}^m), \ldots, g_{q}(V_{Z}^m)]^T =\mathbf{D_1}{\bf v}_q^{m}+\mathbf{D_2}\hat{{\bf g}}_q^{m}$, where $\mathbf{D_1}$ and $\mathbf{D_2}$ are submatrices of $\mathbf{D}$ of dimensions $Z\times(F-Z)$ and $Z\times Z$.
	
	For the subset of shares to leak information, the  key  vector $\hat{{\bf g}}_q^{m}$ must be decoupled from the corresponding secret IV vector ${\bf v}_q^{m}$, i.e., $\mathbf{D_1}{\bf v}_q^{m}\neq\mathbf{0}$ and $\mathbf{D_2}\hat{{\bf g}}_q^{m}=\mathbf{0}$. As all submatrices of a Cauchy matrix are full rank, the columns of $\mathbf{D_2}$ are linearly independent. Hence, such $\hat{{\bf g}}_q^{m}$ does not exist. This implies that a linear combination involving only  ${\bf v}_q^{m}$ cannot be obtained, resulting in zero information leakage from the mapping. Moreover, to demonstrate that an arbitrary node $k$ cannot obtain any information about $\{v_{q,f}^m: q \in [K]\backslash k, f \in [F-Z], m\in [\eta]\}$ from the transmitted messages, observe that each message is encrypted with a key available only to the nodes for which the message is useful. For a transmission $X_{\hat{k},s}^m$ made by some node $\hat{k} \in [K]$, another node $k$ cannot decode any information if  $s$ does not appear in the $k^{\text{th}}$ column of the PDA. This follows since node $k$ does not have any knowledge of $T_{\hat{k},s}^m$. Thus, node $k$ cannot  obtain any information about the linear combination of shares encrypted by this key.

\end{document}